\documentstyle[prl,aps,epsf,floats]{revtex}
\begin{document}
\draft
\twocolumn[\hsize\textwidth\columnwidth\hsize\csname @twocolumnfalse\endcsname
\title{Spectral properties and pseudogaps
       in a model with d-wave pairing symmetry}
\author{Bumsoo Kyung}
\address{D\'{e}partement de physique and Centre de recherche en 
physique du solide.\\
Universit\'{e} de Sherbrooke, Sherbrooke, Qu\'{e}bec, Canada J1K 2R1}
\date{March 30, 2000}
\maketitle
\begin{abstract}
   A model with d-wave pairing symmetry is studied 
by employing a non-perturbative sum rule approach.
At low temperature
the magnitude of a normal state pseudogap 
shows strong $\vec{k}$ or angle dependence
well fitted by $\cos 2\phi$ form.
With increasing temperature, the pseudogap closes at some critical 
angle $\phi_{c}$ and beyond this angle a single quasiparticle-like
peak appears.
The resulting Fermi surface is strongly temperature dependent.
Both in the spectral function and the density of states, the pseudogap
disappears in a manner that the spectral weight fills in 
the pseudogap instead of closing it with increasing temperature.
All these features are qualitatively consistent with ARPES for underdoped
cuprates.
\end{abstract}
\vskip2pc]
\narrowtext

   The nature of an excitation gap in high-temperature superconductors
has been one of the puzzling issues in the community of 
condensed matter physicists.
Through several years of extensive experimental work,
general consensus regarding the superconducting gap symmetry 
seems to be reached that 
the superconducting gap has mainly d-wave character with possibility 
of a small mixture of other angular momentum 
states,\cite{Levi:1993,%
Shen:1993,Marshall:1996}
in contrast to conventional BCS superconductors with an isotropic 
s-wave gap.  
Recent discovery of 
a normal state pseudogap in underdoped cuprates has shed another side 
of the anomalous behaviors in the copper oxide superconductors.
For these materials the low frequency spectral weight begins to 
be strongly suppressed  
below some characteristic temperature 
$T^{*}$ higher than $T_{c}$.
This behavior has been observed through various experimental
probes such as  
photoemission,\cite{Ding:1996,Loeser:1996,Ding:1998}
specific heat,\cite{Loram:1993} 
tunneling,\cite{Renner:1998}
NMR,\cite{Takigawa:1991} and 
optical conductivity.\cite{Orenstein:1990}
In particular
recent angle resolved photoemission spectroscopy
(ARPES)\cite{Ding:1996,Loeser:1996,Ding:1998}
and tunneling experiments\cite{Renner:1998}
indicate that 
the pseudogap phenomenon is closely related to pairing fluctuations.
These measurements
clearly exhibit that
the normal state pseudogap has the same angular dependence
and magnitude as the
superconducting gap and that 
often the only difference between
the spectra in the pseudogap state and the superconducting state
is in their linewidths.
Typically $T^{*}$ is much higher than $T_{c}$ and their doping 
dependence is qualitatively different.
While $T_{c}$ decreases with underdoping, $T^{*}$ increases in contrast.
This feature suggests that $T^{*}$ does not follow 
$T_{c}$ characterized by long-range phase coherence,
but instead some kind of a mean-field critical temperature $T_{MF}$.
In spite of several possible scenarios such as 
the spinon pair formation without 
the Bose-Einstein condensation of holons,\cite{Anderson:1987,Tanamoto:1992,%
Lee:1997}
strong superconducting phase fluctuations%
\cite{Doniach:1990,Emery:1995,Franz:1998} and 
a magnetic scenario near the antiferromagnetic instability\cite{Shen:1997,%
Schmalian:1998} and so on,
at present there is no consensus in the origin of the pseudogap.  

   For the past several years extensive theoretical effort 
has been also made by several groups
to understand
this anomalous pseudogap behavior in the context of short 
range effective (attractive) interaction between electrons.
This may be divided into two 
different classes of approach.
In the first class,
quantum Monte Carlo (QMC) simulations were made for the attractive Hubbard
model.\cite{Scalettar:1989,Moreo:1991,Randeria:1992,Assaad:1993,Trivedi:1995,Singer:1996,Vilk:1998,Allen:1999} 
Although this Hamiltonian is not a realistic model for understanding 
the complex physical behaviors in the underdoped cuprates,
it is believed to capture the important ingredient of 
the paring fluctuations in those materials.
In the absence of a small parameter this numerical method has played 
an important role in understanding of the model, 
in spite of some 
uncertainties due to finite size effect and numerical analytical 
continuation.
In this approach  
the one-particle spectral function as well as 
the density of states clearly show
the precursor of the superconducting gap in the normal state.
In the second class,
various assumptions and approximations are used.
This class includes 
the effect of vortex phase fluctuations on the 
single-particle properties\cite{Franz:1998}
and the `paring approximation' 
theory,\cite{Janko:1997} and 
T-matrix and 
self-consistent T-matrix approximations for the two-dimensional
attractive Hubbard 
model.\cite{Serene:1989,Fresard:1992,Micnas:1995,Deisz:1998,Kyung:1998,%
Kornilovitch:1999}
More recently the self-consistent T-matrix approach was also applied 
to a model with $d_{x^{2}-y^{2}}$ pairing.\cite{Engelbrecht:1998,Hotta:1999}
The present approach is a variant of the T-matrix approximation applied
to a model with d-wave pairing symmetry.

   In a previous study,\cite{Kyung:1999}
a non-perturbative sum rule approach 
was developed for the attractive Hubbard model by extending
previous work on the repulsive Hubbard model.\cite{Vilk:1997}
It is found
that in two dimensions, the mean-field transition temperature is replaced by
a crossover temperature where the correlation length starts to grow
exponentially. At sufficiently low temperature, a Kosterlitz-Thouless O(2)
transition should occur, but it is not reproduced by that approach since it
is in the $O(n=\infty)$ universality class\cite{Dare:1996}. Nevertheless, the
agreement with Monte Carlo calculations is quantitative for both one and
two-particle correlation functions over the whole range of parameters
accessible by Monte Carlo calculations where it is found that in two
dimensions, a pseudogap appears in the single-particle spectral
weight\cite{Vilk:1998}
as well as in the density of states.\cite{Moreo:1991}
Recent dimensional crossover
study by Preosti et al.\cite{Preosti:1999} shows that the pseudogap effect
is basically absent in three dimensions. In weak to intermediate coupling,
the appearance of a pseudogap is traced back to the growing critical
pairing fluctuations in the low-temperature renormalized classical regime of
the low-dimensional system. With increasing temperature, the spectral weight
fills in the pseudogap instead of closing it. Furthermore, the pseudogap
appears earlier in the density of states than in the spectral function.
It was also noted that the qualitative features found in this study should
apply to the d-wave case. In this paper we study in detail spectral
properties and pseudogaps in a model Hamiltonian with d-wave pairing
symmetry which is more appropriate for high-temperature superconductors.

   We consider  
on a two dimensional square lattice 
a simple model Hamiltonian which has a superconducting 
ground state with d-wave symmetry 
\begin{eqnarray}
 H &=& \sum_{\vec{k},\sigma}\varepsilon_{\vec{k}}c^{+}_{\vec{k},\sigma}
      c_{\vec{k},\sigma}
                                             \nonumber  \\
   &+&\frac{1}{N}\sum_{\vec{k},\vec{k}',\vec{q}}
      V_{\vec{k},\vec{k}'}
      c^{+}_{\vec{k},\uparrow}
      c^{+}_{-\vec{k}+\vec{q},\downarrow}
      c_{-\vec{k}'+\vec{q},\downarrow} 
      c_{\vec{k}',\uparrow}  \; ,
                                                           \label{eq1}
\end{eqnarray}
where $\varepsilon_{\vec{k}}=-2t(\cos k_{x}+\cos k_{y})$
and $N$ is the number of
lattice sites.
We choose 
$V_{\vec{k},\vec{k}'}$ as a d-wave separable potential given as 
$V\Pi(\vec{k})\Pi(\vec{k}')$ where
$\Pi(\vec{k})=\cos k_{x}-\cos k_{y}$, or as 
 $ 1.5264\cos 2\phi
  =1.5264 \frac{k_{x}^{2}-k_{y}^{2}}
              {k_{x}^{2}+k_{y}^{2}}$ where 
$\phi=\arctan \frac{k_{y}}{k_{x}}$.
In the real space notation the interaction term may be written as 
$H_{\mbox{I}}=V\sum_{i} \Delta^{+}_{i} \Delta_{i}$ where 
$\Delta_{i}=\sum_{\delta}g(\delta)c_{i,\downarrow}
                                  c_{i+\delta,\uparrow}$.
$g(\delta)$ is the Fourier transform of $\Pi(\vec{k})$ given as 
$g(\delta)=\frac{1}{N}\sum_{\vec{k}} e^{i\vec{k} \cdot \vec{\delta}}
 \Pi(\vec{k})$.
For example, for the following pair structure, 
\[ g(\delta) = \left\{ \begin{array}{lll}
                          1/2     &  \mbox{if $\delta=(\pm 1,0)$} , \\
                         -1/2     &  \mbox{if $\delta=(0,\pm 1)$} , \\
                           0      &  \mbox{if otherwise} ,
                       \end{array}
               \right.  \]
it leads to 
$\Pi(\vec{k})=\cos k_{x}-\cos k_{y}$.
Obviously  
$\Pi(\vec{k})=1.5264\cos 2\phi$
requires in $g(\delta)$ further neighbors beyond the four nearest 
neighbors.
Here a numerical factor of 1.5264 is introduced to normalize  
$g(\delta)$ with
$\sum_{\delta} g^{2}(\delta)=1$.
In terms of $\Delta_{i}$ the d-wave pairing susceptibility is defined as
\begin{eqnarray}
\chi_{pp}(i,\tau) = \langle T_{\tau} \Delta_{i}(\tau) 
                                     \Delta^{+}_{0} \rangle  \; ,
                                                           \label{eq2}
\end{eqnarray}
where $T_{\tau}$ is the imaginary time ordering operator. 

   It is well-known that
for many strongly correlated electronic models
there is no obvious small parameter with which
a systematic perturbative approximation can be made.
In such a situation,
a non-perturbative approach may be a good alternative for at least
qualitative understanding of
the physics in those strongly correlated model Hamiltonians.
In this paper
we apply a non-perturbative sum rule approach to
the model Hamiltonian~(\ref{eq1}),
which is still not a fully controlled approximation in nature.
Because of an extended nature of $\Delta_{i}$ for a d-wave pair
and the increasing difficulty of the derivation in the real space 
representation, here we use the analogy from our previous studies for 
the repulsive\cite{Vilk:1994,Vilk:1997} 
and attractive Hubbard models.\cite{Kyung:1999}
These studies show the important many-body modification
with respect to the standard RPA and T-matrix approaches.
The modification comes in two different places.
First, the paring susceptibility is calculated through vertex function
$U_{pp}$ instead of bare interaction strength $U$, which is constant 
in our approximation.
Second, the pairing fluctuations are coupled to electrons, leading to the 
self-energy
with $UU_{pp}$ form instead of $U^{2}$. 
In particular, the latter structure
is in agreement with the fact that there is no Migdal
theorem for this problem, contrary to the case of electron-phonon
interactions.
And the 
renormalized interaction constant $U_{pp}$ is determined by 
the exact sum rule for the pairing susceptibility.
By using the sum rule and the renormalized constant $U_{pp}$, 
we determine in effect the Ginzburg-Landau parameters due to mode-mode
coupling. Thus this approach is similar to the one-loop renormalization 
group approach within the Gaussian approximation and as a result the 
Mermin-Wager theorem is formally satisfied
in two dimensions.\cite{Mermin:1966}
This sum rule approach was systematically compared with 
the QMC simulations for the 
repulsive and attractive Hubbard models and the agreement 
was in a quantitative level
both in one- and two-particle functions.
We believe that this generic feature of the many-body modification  
should also carry over 
to a Hamiltonian with d-wave pairing symmetry.
In this modification, the pairing susceptibility and the self-energy
can be written in Fourier space
in terms of renormalized interaction strength
$V_{pp}$
\begin{eqnarray}
\chi_{pp}(q) &=& \frac{ \chi^{0}_{pp}(q)}{1+V_{pp}\chi^{0}_{pp}(q)} \; ,
                                             \nonumber  \\
\Sigma(k) &=& - VV_{pp}\Pi^{2}(\vec{k})\frac{T}{N}\sum_{q}
                \chi_{pp}(q)
                G^{0}(q-k)
                                               \; ,
                                                           \label{eq3}
\end{eqnarray}
where the irreducible susceptibility is defined as 
\begin{eqnarray}
\chi^{0}_{pp}(q) = \frac{T}{N}\sum_{k}\Pi^{2}(\vec{k})G^{0}(q-k)G^{0}(k)
                                               \; .
                                                           \label{eq4}
\end{eqnarray}
The above expressions are reduced to 
the standard T-matrix approximation when $V_{pp}$ is replaced 
by bare $V$. 
We determine this renormalized constant by employing the exact
sum rule for the d-wave pairing susceptibility
\begin{eqnarray}
& &\frac{T}{N}\sum_{q}\chi_{pp}(q)e^{-i\nu_{m}0^{-}}  
                                             \nonumber  \\
&=&  \sum_{\delta,\delta'} g(\delta) g(\delta')
    \langle  c^{+}_{\delta',\downarrow}
      c_{\delta,\downarrow} 
      c^{+}_{\uparrow}
      c_{\uparrow} \rangle
                                               \; .
                                                           \label{eq5}
\end{eqnarray}
Note that the convergence factor is necessary in the sum rule, because  
away from half-filling $\chi_{pp}(i\nu_{m})$ decays like 
$1/\nu_{m}$ at large frequencies.  
In the previous studies for the repulsive and attractive Hubbard models,
the right-hand side of the sum rule 
evaluated in the SDW and BCS mean-field ground states, respectively, was 
found to be in excellent agreement with the QMC values in the 
intermediate to strong coupling regimes.\cite{Comment1}
In this paper the right-hand side of Eq.~(\ref{eq5})
is also evaluated in the d-wave BCS mean-field ground state 
\begin{eqnarray}
&&    \frac{1}{N^{2}}\sum_{\vec{k},\vec{p},\vec{k}',\vec{p}'}
     \Pi(\vec{k})\Pi(\vec{p})
     \langle 
      c^{+}_{\vec{k},\downarrow}
      c_{\vec{p},\downarrow} 
      c^{+}_{\vec{k}',\uparrow}
      c_{\vec{p}',\uparrow} \rangle \delta_{\vec{k}+\vec{k}',
                                    \vec{p}+\vec{p}'} 
                                             \nonumber  \\
& = &  [\frac{1}{N}\sum_{\vec{k}}
     \Pi^{2}(\vec{k})v^{2}(\vec{k})]
     [\frac{1}{N}\sum_{\vec{k}}
     v^{2}(\vec{k})]
                                             \nonumber  \\
& + &  [\frac{1}{N}\sum_{\vec{k}}
     \Pi(\vec{k})u(\vec{k})v(\vec{k})]^{2}
                                               \; ,
                                                           \label{eq6}
\end{eqnarray}
where 
\begin{eqnarray}
 u_{\vec{k}}^{2} & = &
   \frac{1}{2}(1+\frac{\varepsilon_{\vec{k}}-\mu}{E_{\vec{k}}}) \; ,
                                                         \nonumber  \\
 v_{\vec{k}}^{2} & = &
   \frac{1}{2}(1-\frac{\varepsilon_{\vec{k}}-\mu}{E_{\vec{k}}}) \; ,
                                                         \nonumber  \\
 E_{\vec{k}} & = & \sqrt{
   (\varepsilon_{\vec{k}}-\mu)^{2}+\Delta^{2}(\vec{k})}  \; .
                                                           \label{eq7}
\end{eqnarray}
Here $\Delta(\vec{k})$ is the d-wave BCS mean-field gap. 
The chemical potential $\mu$ and the gap $\Delta(\vec{k})$ are determined
self-consistently through the number and gap equations
for given $V$, $T$ and $n$.

   Before starting we comment on some differences 
associated with $\Pi(\vec{k})=\cos k_{x}-\cos k_{y}$ and 
with $\Pi(\vec{k})=1.5264\cos 2\phi$ structure.
The interaction Hamiltonian with the first structure  
depends not only on the angle 
of $\vec{k}$
but also on its magnitude.
Thus the pairing interaction is always strongest 
at $\vec{k},\vec{k}'=(\pm \pi,0)$ or $(0,\pm \pi)$, although 
the noninteracting Fermi surface can be far from 
these points.
As a result two features occur for a particle density  
far away from half-filling (Detailed 
calculations were performed but not shown in this paper).
First, the peaks associated with  
the precursor of the Bogoliubov quasiparticles 
occur asymmetrically with respect to 
the Fermi energy. 
Second, the locus of $\vec{k}$ points satisfying 
$\omega-\varepsilon_{\vec{k}}+\mu-Re\Sigma(\vec{k},\omega)=0$ at 
$\omega=0$
can be substantially different from the noninteracting Fermi surface,
thus strongly violating the Luttinger's theorem. 
For $\Pi(\vec{k})=1.5264\cos 2\phi$
which depends only on the angle, however,
the above features disappear and 
the locus of $\vec{k}$ points satisfying 
the above equation
is almost identical with the noninteracting Fermi surface. 
Near half-filling the differences are negligible.
Throughout the calculations,
lattice spacing, $\hbar$, and $k_{B}$ are set to be unity, and 
all energies are measured in unit of $t$.
We used a discrete lattice as large as 
$128 \times 128$ in momentum space
and performed the calculations by means of
fast Fourier transforms (FFT).
Equations (\ref{eq3}) and (\ref{eq4}) are computed
in terms of Matsubara frequencies
and the analytic continuation from Matsubara to real frequencies
are made via Pade approximants.\cite{Vidberg:1977}
In order to detect any spurious features
associated with numerical analytical continuation,
we performed real frequency calculations in parallel.
Except for 
some spiky features 
in the real frequency formulation
coming from the Lorentzian approximation of the
non-interacting Green's function, 
the two results are almost identical.

\begin{figure}
 \vbox to 7.0cm {\vss\hbox to -5.0cm
 {\hss\
       {\includegraphics{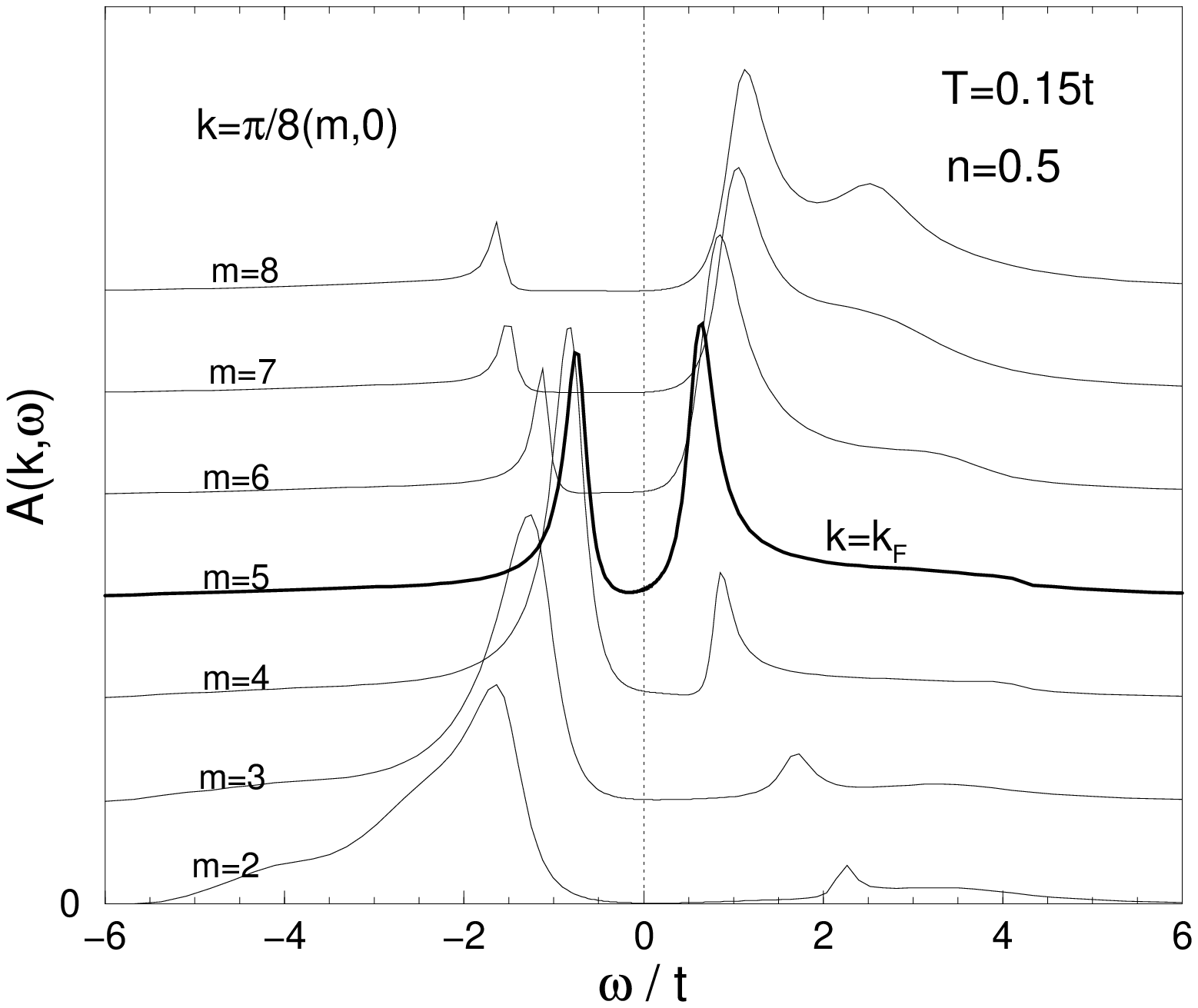}
       }
  \hss}
 }
\caption{Spectral function along the line $(0,0)-(\pi,0)$
         for $V=-4$, $n=0.5$ and $T=0.15$.
         The figures for $m=3-8$ are shifted vertically
         by 0.25. $m=5$ corresponds to the noninteracting Fermi surface.}
\label{fig1}
\end{figure}

   We begin by presenting the spectral functions along $(0,0)-(\pi,0)$
direction for $V=-4t$, $n=0.5$ and $T=0.15t$ in Fig.~\ref{fig1}.
Throughout the paper the density is fixed at $n=0.5$. The 
results for other densities are similar. 
For this parameter, $V_{pp}$ satisfying the sum rule
is found to be $-1.87t$
significantly different from the bare value ($V=-4t$).
This shows the importance of the mode-mode coupling effect already in the 
intermediate coupling regime.
Let us label wave vectors by $(m\pi/8,0)$.
Below the Fermi wave vector ($m=5$), the main peak stays below 
the Fermi energy and the secondary peak grows in strength as $\vec{k}$
approaches $\vec{k}_{F}$.
At the Fermi wave vector
two peaks reminiscent of the Bogoliubov 
quasiparticles appear almost symmetrically with respect to the 
Fermi energy.
Since electrons are still in the normal state, this is 
the precursor of the superconducting gap, namely, a normal state pseudogap.
Above the Fermi wave vector, the main peak stays above
the Fermi energy and the secondary peak becomes stronger as $\vec{k}$
approaches $\vec{k}_{F}$.  
This result should be contrasted with that by 
Engelbrecht {\it et al.}\cite{Engelbrecht:1998}
In the self-consistent T-matrix approximation with  
$\Pi(\vec{k})=\cos k_{x}-\cos k_{y}$ form, these authors
argued that along $(0,0)-(\pi,0)$ direction
the dominant peak of $A(\vec{k},\omega)$
never crosses the Fermi energy and bounces back towards the negative
frequency.   
Their overall finding, however, is qualitatively different from  
our results.
In the present calculations,
the dominant peak of $A(\vec{k},\omega)$ eventually passes through the 
Fermi energy for $\vec{k}$ far above the Fermi wave vector.
At the noninteracting Fermi surface our spectral weight is strongly suppressed
at the Fermi energy to become a local minimum, while the local 
minimum of $A(\vec{k},\omega)$ in their study is located significantly 
away from the Fermi energy.
Presumably this is due to 
the features 
associated with $\Pi(\vec{k})=\cos k_{x}-\cos k_{y}$ structure 
mentioned in the previous paragraph
and also due to
the self-consistent approximation 
that does not take the vertex function and the Green's 
function at the same level of approximation.\cite{Comment2}

\begin{figure}
 \vbox to 7.0cm {\vss\hbox to -5.0cm
 {\hss\
       {\includegraphics{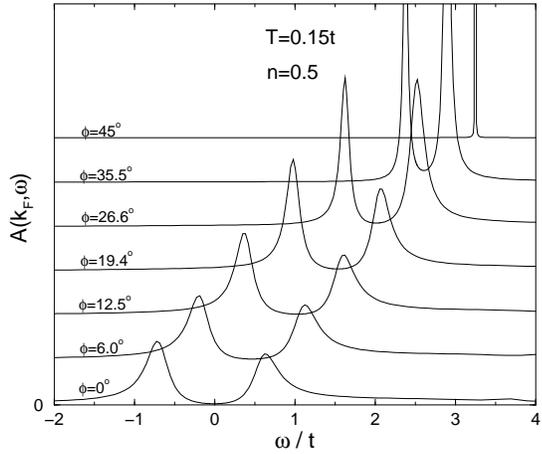}
       }
  \hss}
 }
\caption{Spectral function along the noninteracting Fermi surface
         for $V=-4$, $n=0.5$ and $T=0.15$.
         The angle $\phi$ is defined as $\arctan (k_{y}/k_{x})$ along
         the noninteracting Fermi surface.
         The figures for $\phi > 0^{o}$ are shifted vertically
         and horizontally both by 0.5. }
\label{fig2}
\end{figure}

\begin{figure}
 \vbox to 7.0cm {\vss\hbox to -5.0cm
 {\hss\
       {\includegraphics{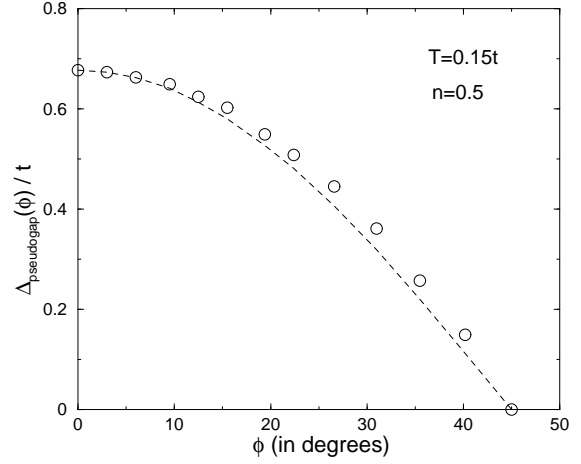}
       }
  \hss}
 }
\caption{The magnitude of the pseudogap
         for $V=-4$, $n=0.5$ and $T=0.150$ (open circles).
         The angle $\phi$ is defined as $\arctan (k_{y}/k_{x})$ along
         the noninteracting Fermi surface.
         The dashed curve is $\Delta(\phi=0^{o})\cos 2\phi$.
         }
\label{fig3}
\end{figure}

   In Fig.~\ref{fig2} the spectral function is 
shown along the Fermi wave vectors at low temperature ($T=0.15t$).
The angle $\phi$ is defined as $\arctan (k_{y}/k_{x})$ along the 
noninteracting Fermi surface.
At $\phi=0^{o}$ the magnitude of the pseudogap is largest and 
as $\phi$ increases it progressively decreases, leading to 
strong momentum or angle dependence in the size of the pseudogap.
Since along the diagonal directions the pairing interaction vanishes,
the pseudogap completely closes and 
a sharp quasiparticle peak appears
at $\phi=45^{o}$. 
In Fig.~\ref{fig3} the magnitude of the pseudogap 
(circles)
is plotted as a function of angle $\phi$ for $T=0.15t$ along 
with the ground state gap symmetry $\Delta(\phi=0^{o})\cos 2\phi$ 
(dashed curve).
The angle dependence of the normal state pseudogap is well fitted
by d-wave symmetry, consistent with ARPES experiments for underdoped cuprates.

\begin{figure}
 \vbox to 7.0cm {\vss\hbox to -5.0cm
 {\hss\
       {\includegraphics{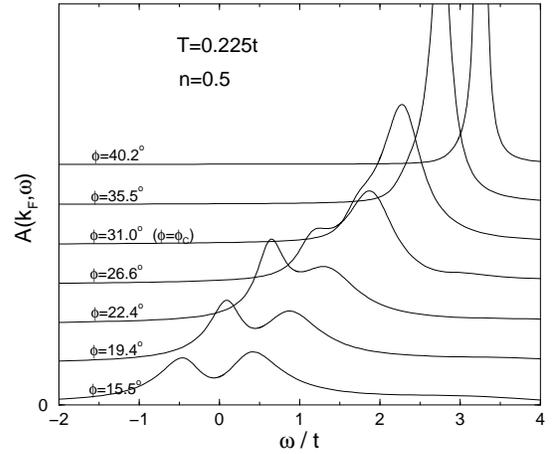}
       }
  \hss}
 }
\caption{Spectral function along the noninteracting Fermi surface
         for $V=-4$, $n=0.5$ and $T=0.225$.
         The angle $\phi$ is defined as $\arctan (k_{y}/k_{x})$ along
         the noninteracting Fermi surface.
         The figures for $\phi > 15.5^{o}$ are shifted vertically
         and horizontally by 0.25 and 0.5, respectively. }
\label{fig4}
\end{figure}

\begin{figure}
 \vbox to 7.0cm {\vss\hbox to -5.0cm
 {\hss\
       {\includegraphics{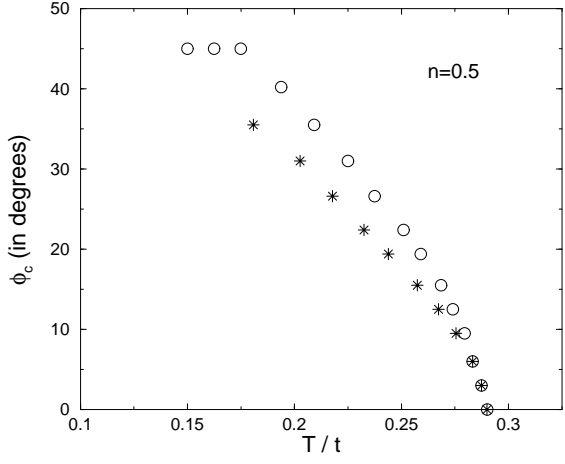}
       }
  \hss}
 }
\caption{Critical angle $\phi_{c}$ as a function of
         temperature for $V=-4$, $n=0.5$ (open circles).
         The stars are the critical angle estimated by the expression
         given in the text.}
\label{fig5}
\end{figure}

    At higher temperatures a drastic change is found in the 
spectral function.
For $T=0.225t$ the pseudogap closes well below 
$45^{o}$, as shown in Fig.~\ref{fig4}.
The local minimum at the Fermi energy disappears
at $\phi=31.0^{o}$ ($26.6^{o}$ is just on the crossover).
Beyond this angle, the quasiparticle-like peak appears and thereby 
the Fermi surface is partially restored.
We define this angle as critical angle $\phi_{c}$.
In Fig.~\ref{fig5} $\phi_{c}$ 
is plotted for different temperatures (open circles).
Below $T=0.175t$ the pseudogap is found everywhere 
along the Fermi wave vectors
except at $\phi=45^{o}$, thus the Fermi surface is destroyed everywhere
except along the diagonal directions.
With increasing temperature (up to $T=0.29t$), however,
$\phi_{c}$ becomes smaller than $45^{o}$.
Thus, the pseudogap region shrinks and at the same time
the Fermi surface grows from $\phi=45^{o}$ 
up to the critical angle. 
Above $T=0.29t$ the whole Fermi surface is completely restored
in spite of some broadening of the spectral function  
due to interaction as 
well as finite temperature.
This feature is qualitatively consistent with ARPES for underdoped 
cuprates.\cite{Ding:1998}
We can theoretically calculate the temperature dependence of
$\phi_{c}$.
By using the Ornstein-Zernike form of the pairing correlation
function and taking the classical fluctuations ($iq_{n}=0$),\cite{Vilk:1996} 
the scattering rate at the Fermi energy is found to be proportional to 
$\Pi^{2}(\vec{k})\xi/\xi_{T}$, where $\xi$ and $\xi_{T}$ are 
pairing correlation length and thermal de Broglie wave length
defined as $\xi_{T}=v_{F}(\vec{k})/T$, respectively.
For the pseudogap to disappear, the scattering rates should be much smaller
than unity
$\Pi^{2}(\vec{k})\xi/\xi_{T} \ll 1$,
allowing us to define a critical angle
$\xi = v_{F}(\phi_{c})/T/\Pi^{2}(\phi_{c})$.
In Fig.~\ref{fig5} this critical angle is also shown as stars.
For the best fit near $0^{o}$ angle, a numerical factor of 1.06
multiplies $\xi$.
Although a small deviation is found near $45^{o}$, the overall magnitude and 
shape are in reasonable agreement throughout the whole angle.
For a d-wave model, the condition for the appearance of a pseudogap 
in a given momentum
depends not only on the anisotropy of the Fermi velocity 
(which is the only relevant condition in the attractive Hubbard model)
but also more importantly on an angle dependent form factor
$\Pi^{2}(\vec{k})$. 

\begin{figure}
 \vbox to 7.0cm {\vss\hbox to -5.0cm
 {\hss\
       {\includegraphics{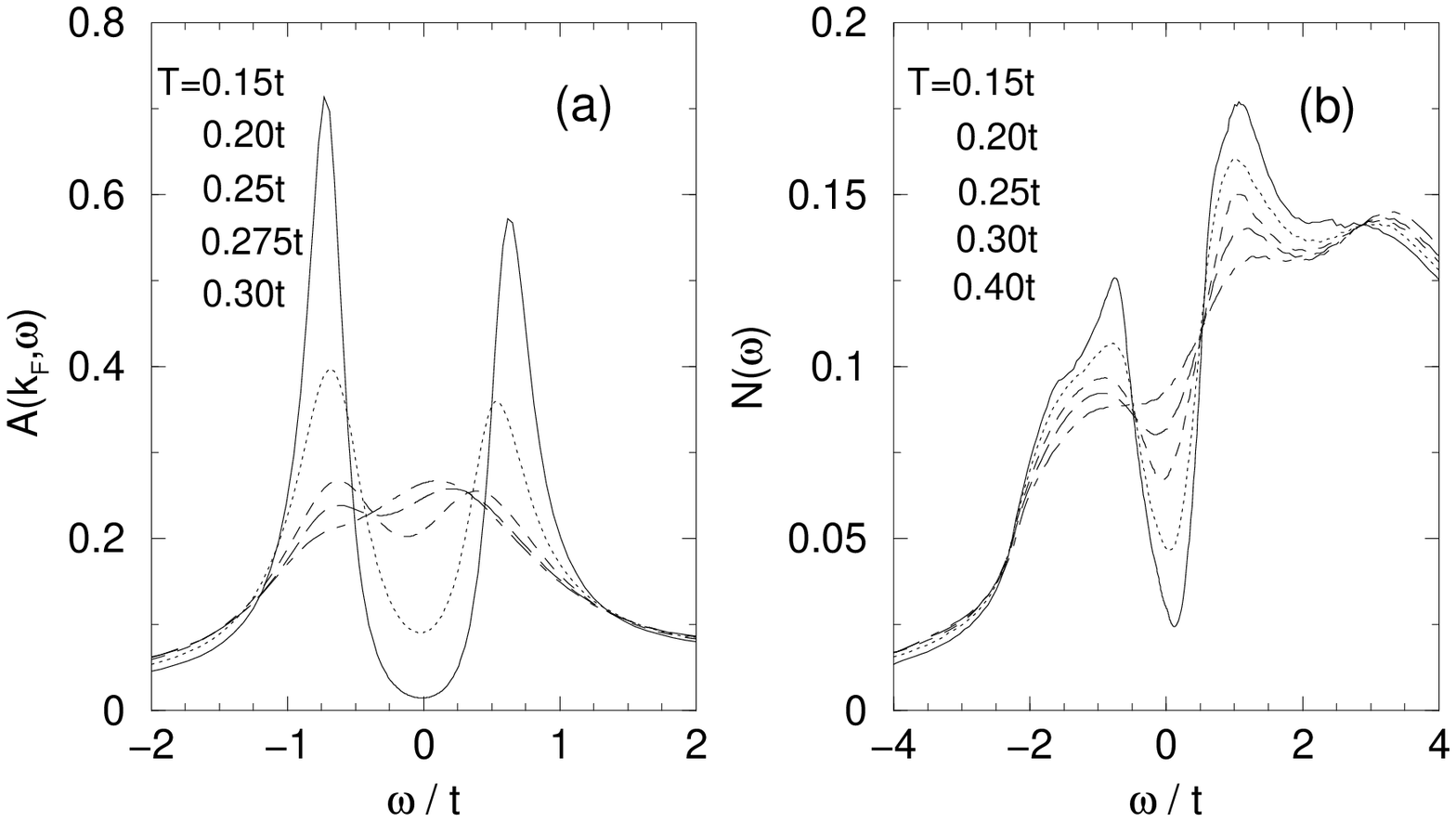}
       }
  \hss}
 }
\caption{(a) The spectral function at the noninteracting Fermi surface and
         (b) the density of states
             for $V=-4$ and $n=0.5$ at different temperatures.
             The solid, dotted, dashed, long-dashed, and dot-dashed curves
             correspond to temperatures from top to bottom.}
\label{fig6}
\end{figure}

   In Fig.~\ref{fig6}
we show the spectral function at the Fermi wave vector and 
the density of states for different temperatures.
As the temperature is increased, the spectral weight starts to fill 
in the pseudogap and at $T=0.3t$ the precursor of the superconducting 
gap completely disappears as shown in Fig.~\ref{fig6}(a).
At this temperature, however, the pseudogap still persists
in the density of states and at higher temperature ($T=0.4t$)
it finally disappears as shown
in Fig.~\ref{fig6}(b).
Except close to half-filling, the pseudogap appears at higher 
temperature in the density of states than in the spectral function.
Compared with the s-wave case, the density of states is suppressed
linearly near the Fermi energy, a reminiscence of 
$N(\omega) \sim \omega$ in the superconducting state.
Both in the spectral function and the density of states, the pseudogap
disappears in a manner that the spectral weight fills in 
the pseudogap instead of closing it with increasing temperature.
This feature is also consistent with ARPES for underdoped cuprates.
This may suggest that in our approach
phase fluctuations (spin-wave type) rather than
amplitude fluctuations are mainly responsible for the pseudogap formation,
although the present approach includes both. 
Like in the s-wave case,
a pseudogap also appears in the density of 
states when the characteristic pairing frequency
$\nu_{c}$ is equal to or smaller than temperature.
This corroborates the origin of the d-wave pseudogap, namely,
growing d-wave paring fluctuations in the low temperature
renormalized classical regime of
the low dimensional system.
The calculated $T^{*}$ follows the same trend as the mean-field critical 
temperature $T_{MF}$. (To be more precise, $T^{*}$ is approximately half of 
$T_{MF}$ for most of the densities.)
As noted in Ref.\cite{Allen:1999}, near a point with high order 
parameter symmetry (half-filling in that paper), 
the transition temperature $T_{c}$ decreases
while the pseudogap temperature increases along with $T_{MF}$.
As a result it leads to a large pseudogap regime, consistent with  
the phase diagram  
in the underdoped side of cuprates.

   There are several advantages
in our formulation.
First, the pairing fluctuation sum rule Eq.~(\ref{eq5}) 
is exactly satisfied (by construction).
Through this sum rule, the Mermin-Wagner theorem is formally fulfilled and 
the strength of pairing fluctuations is properly constrained within the 
Gaussian approximation.
This latter feature 
is crucial in our formulation, because
an approximate treatment of pairing fluctuations without 
constraining the strength
can easily overestimate or 
underestimate the magnitude of fluctuations particularly
in low dimensional systems.
The Mermin-Wagner theorem is also satisfied in some other approaches
such as the self-consistent T-matrix (FLEX) and
the ``pairing'' approximation schemes.
Second, there is an exact relation between
one-particle (self-energy, Green's function)
and two-particle (interaction term) functions:
\begin{eqnarray}
\lim_{\tau \rightarrow 0^{-}} \frac{T}{N} \sum_{\vec{k},i\omega_{n}}
     \Sigma(\vec{k},i\omega_{n})
     G(\vec{k},i\omega_{n})
     e^{-i\omega_{n}\tau}
    =V \langle \Delta^{+}_{i}\Delta_{i} \rangle \; ,
                                                           \label{eq8}
\end{eqnarray}
where here the self-energy includes the Hartree-Fock term.
When the interacting Green's function  
$G(\vec{k},i\omega_{n})$ is replaced by the noninteracting one 
$G^{0}(\vec{k},i\omega_{n})$ in Eq.~(\ref{eq8}), it is  
exactly satisfied. 
With $G(\vec{k},i\omega_{n})$, the difference between the left- and 
right-hand sides is less than $6\%$ for all temperatures studied.
In this paper the Kosterlitz-Thouless phase transition\cite{Kosterlitz:1973}
and its fluctuation effect near $T_{KT}=T_{c}$ have been neglected, since  
the present formulation is inadequate to describe the topological nature of
the phase in two dimensions and vortex anti-vortex binding-unbinding 
physics.
Since for similar parameters for the attractive Hubbard model the QMC 
results indicate that 
$T_{KT} \sim 0.05t$\cite{Moreo:1991} three times smaller than the lowest  
temperature in our calculations,
we do not expect in the present results any significant influence 
from vortex phase fluctuations. 
Our approach is valid for weak to intermediate coupling and for 
temperature not too deep in the pseudogap regime. At very low temperature,
even the Ginzburg-Landau functional form itself
may change, for instance, 
a possible crossover of the dynamical critical exponent 
from z=2 to other value,\cite{Sachdev:1995}
and eventually vortex phase fluctuations may come into 
play.

   In summary,
we have studied spectral properties and pseudogaps
in a model with d-wave pairing symmetry by using a non-perturbative 
sum rule approach.
We applied to this model our previous 
experience of many-body theory in the repulsive and 
attractive Hubbard models.
The magnitude of the normal state pseudogap shows 
strong angle dependence
well fitted by $\cos 2\phi$ form at low temperature.
With increasing temperature the pseudogap closes at some critical 
angle $\phi_{c}$ and beyond this angle a single quasiparticle-like
peak appears.
The resulting Fermi surface is strongly temperature dependent.
With increasing temperature, the pseudogap region shrinks and 
at the same time the Fermi arc grows
from $\phi=45^{o}$ to 
$\phi_{c}$.
Both in the spectral function and density of states the pseudogap
disappears in a manner that the spectral weight fills in 
the pseudogap instead of closing it with increasing temperature.
All these features are qualitatively consistent with ARPES for underdoped
cuprates.
The pseudogap is caused by growing d-wave critical pairing 
fluctuations in the low-temperature classical renormalized regime
of the low-dimensional system, as in the repulsive 
and attractive Hubbard models.
We argue that although the real critical behaviors and critical exponents
are governed by the vortex phase fluctuations close to the $T_{KT}$,
the initial growth of pairing fluctuations can be driven by spin-wave 
phase fluctuations, leading to  
the normal state pseudogap formation.
The behavior of the spin-wave type phase fluctuations belonging to 
the $O(n=\infty)$ universality class\cite{Dare:1996} 
can be qualitatively different in two dimensions (particularly  
near a point with high order 
parameter symmetry) 
from that of   
the $O(2)$ Kosterlitz-Thouless vortex phase fluctuations.

   The author would like to thank A. M. Tremblay for numerous help and
discussions throughout the work.
This work was supported by a grant from the Natural Sciences and
Engineering Research Council (NSERC) of Canada and the Fonds pour la
formation de Chercheurs et d'Aide \`a la Recherche (FCAR) of the Qu\'ebec
government.
\newpage

\begin{references}
\bibitem{Comment1}
In the weak coupling limit $(|U| \leq t-2t)$, 
however, the right-hand side (double occupancy
in the Hubbard model) evaluated in the BCS ground state is 
underestimated compared with the exact value. This is in contrast to a general 
belief that in the weak coupling limit the BCS ground state is exact.
Through an interpolation scheme,
a weak coupling case may be covered as well.
In this paper the intermediate coupling 
regime is studied.
\end{references}
\end{document}